\documentclass[prb,showpacs,twocolumn,floatfix,superscriptaddress]{revtex4-1}
\usepackage[version=3]{mhchem} 
\usepackage{upgreek} 
\usepackage{amsmath}    
\usepackage{graphicx}   
\usepackage{verbatim}   
\usepackage{color}      
\usepackage{subfigure}  
\usepackage{hyperref}   
\begin{document}

\title{Three megahertz photon collection rate from an \\NV center with millisecond spin coherence}

\author{Luozhou Li}
\affiliation{These authors contributed equally.}
\affiliation{Department of Electrical Engineering and Computer Science, Massachusetts Institute of Technology, Cambridge, Massachusetts 02139, United States}

\author{Edward H. Chen}
\affiliation{These authors contributed equally.}
\affiliation{Department of Electrical Engineering and Computer Science, Massachusetts Institute of Technology, Cambridge, Massachusetts 02139, United States}

\author{Jiabao Zheng}
\affiliation{Dept. of Electrical Engineering, Columbia University, New York, NY 10027, United States}

\author{Sara L. Mouradian}
\affiliation{Department of Electrical Engineering and Computer Science, Massachusetts Institute of Technology, Cambridge, Massachusetts 02139, United States}

\author{\\Florian Dolde}
\affiliation{Department of Electrical Engineering and Computer Science, Massachusetts Institute of Technology, Cambridge, Massachusetts 02139, United States}

\author{Tim Schr\"{o}der}
\affiliation{Department of Electrical Engineering and Computer Science, Massachusetts Institute of Technology, Cambridge, Massachusetts 02139, United States}

\author{Sinan Karaveli}
\affiliation{Department of Electrical Engineering and Computer Science, Massachusetts Institute of Technology, Cambridge, Massachusetts 02139, United States}

\author{Matthew L. Markham}
\affiliation{Element Six, 3901 Burton Drive, Santa Clara, CA 95054, USA}

\author{\\Daniel J. Twitchen}
\affiliation{Element Six, 3901 Burton Drive, Santa Clara, CA 95054, USA}

\author{Dirk Englund}
\email{englund@mit.edu}
\affiliation{Department of Electrical Engineering and Computer Science, Massachusetts Institute of Technology, Cambridge, Massachusetts 02139, United States}

\date{11 Sept 2014}

\begin{abstract}
Efficient collection of the broadband fluorescence of the diamond nitrogen vacancy center is essential for a range of applications in sensing, on-demand single photon generation, and quantum information processing. Here, we introduce a circular `bullseye' diamond grating enabling a collected photon rate of $(3.0\pm0.1)\times10^6$ counts per second from a single nitrogen-vacancy center with a spin coherence time of 1.7$\pm$0.1~ms. Back-focal-plane studies indicate efficient redistribution into low-NA modes.
\end{abstract}

\maketitle


The exceptional optical and spin properties in diamond\cite{doherty2013nitrogen,jelezko_single_2006} has led to the demonstration of a wide range of quantum technologies including quantum entanglement\cite{togan2010quantum, bernien2013heralded,dolde_room-temperature_2013}, teleportation\cite{pfaff2014unconditional}, and sensing\cite{2008.NPhys.Taylor, dolde2011electric, kucsko2013nanometre,sushkov_all-optical_2013}. Central to all of these experimental efforts is the efficient detection of the NV photoluminescence (PL), which improves the sensitivity in metrology applications\cite{rondin2014magnetometry} and allows for faster quantum information processing\cite{bernien2013heralded, waldherr2014quantum, taminiau2014universal, blok_manipulating_2014}. However, efficient photon collection has been hindered by total internal reflection confinement due to the high refractive index of diamond. Previous approaches to address this problem in bulk materials include solid immersion lenses\cite{hadden2010strongly, marseglia2011nanofabricated, schroder2011ultrabright, pfaff2014unconditional} ($1.1\times10^6$~cts/s reported), vertical pillars\cite{2009.NNano.Loncar.diamond_nanowire, neu2014photonic,momenzadeh_nano-engineered_2014} ($1.5\times10^6$ cts/s), optical antennas\cite{riedel2014diamond} ($0.6\times10^6$ cts/s) and silicon dioxide gratings\cite{choy2013spontaneous} ($0.7\times10^6$ cts/s). Here, we introduce a circular diamond `bullseye' grating that achieves the highest reported photon collection rate from a single NV center of $(3.0\pm0.1)\times10^6$~cts/s. We measure a spin coherence time of 1.7$\pm$0.1~ms, comparable to the highest reported spin coherence times of NVs under ambient conditions\cite{bar2013solid,naydenov_dynamical_2011,balasubramanian_ultralong_2009}. The planar architecture allows for on-chip integration, and the circular symmetry supports left- and right-handed circularly polarized light for spin-photon entanglement\cite{togan2010quantum}.

\begin{figure} 
	\centering
		\includegraphics[width=0.5\textwidth]{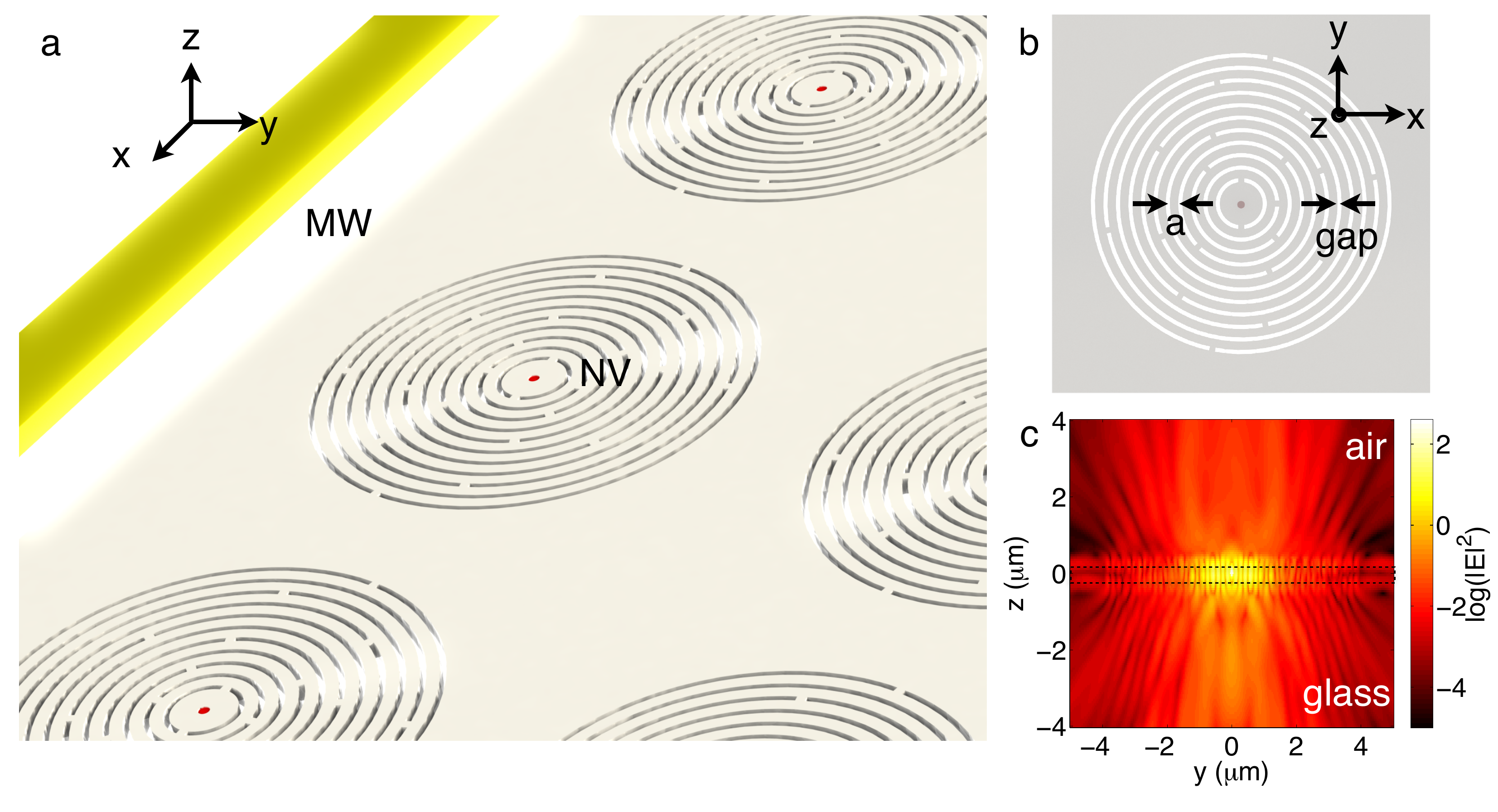}
	\caption{
	(a) Illustration of an array of diamond bullseye gratings adjacent to a microwave (MW) strip line. 
	(b) Schematic of the circular grating. $a$ denotes the lattice constant and $gap$ the air spacing between circular gratings. 
	(c) Energy density (log scale) in the x~=~0 plane with air above and glass below the diamond. An oil objective lens was used for excitation and collection from the bottom. }
\label{fig: 1}
\end{figure}

The bullseye grating consists of concentric slits fully etched into a diamond membrane (Fig.~1a). The grating period $a$ satisfies the second-order Bragg condition, $a = \lambda/n_{eff}$, where $\lambda\sim680$~nm approximates the mean of the NV emission wavelength and $n_{eff}$ is the membrane's effective index when placed on glass\cite{ates2012bright}. Figure~1c shows the simulated field distribution of the bullseye grating with period of $a$~=~330~nm, and an air gap of 99~nm (Lumerical, FDTD Solutions). Light guided in the membrane scatters with equal phase at the slits, leading to constructive interference in the vertical direction. As seen in Fig.~1c, PL from a dipole emitter oriented in the crystallographic direction (54$^\circ$ from vertical) is preferentially ($\sim70\%$) emitted into the glass coverslip due to a lower index contrast of the diamond-glass compared to the diamond-air interface.

The diamond structure was fabricated by first thinning $\sim$5~$\upmu$m-thick diamond membranes to $\sim$300~nm in a reactive ion etcher. The diamond was grown by microwave plasma assisted chemical vapor deposition and contained a density of intrinsic NVs of $\sim$1/$\upmu$m$^3$ and a nitrogen concentration of <100~ppb. The grating patterns were transferred into the diamond membranes using pre-patterned single-crystal silicon membranes as etch masks. These silicon membrane hard masks were positioned onto the diamond and mechanically removed after etching\cite{li2014coherent}. Fig.~2a shows a scanning electron micrograph of a typical fabricated structure. These membranes were subsequently transferred onto a glass coverslip with a pre-patterned microwave strip line for optical and spin characterization. 

\begin{figure} 
	\centering
		\includegraphics[width=0.5\textwidth]{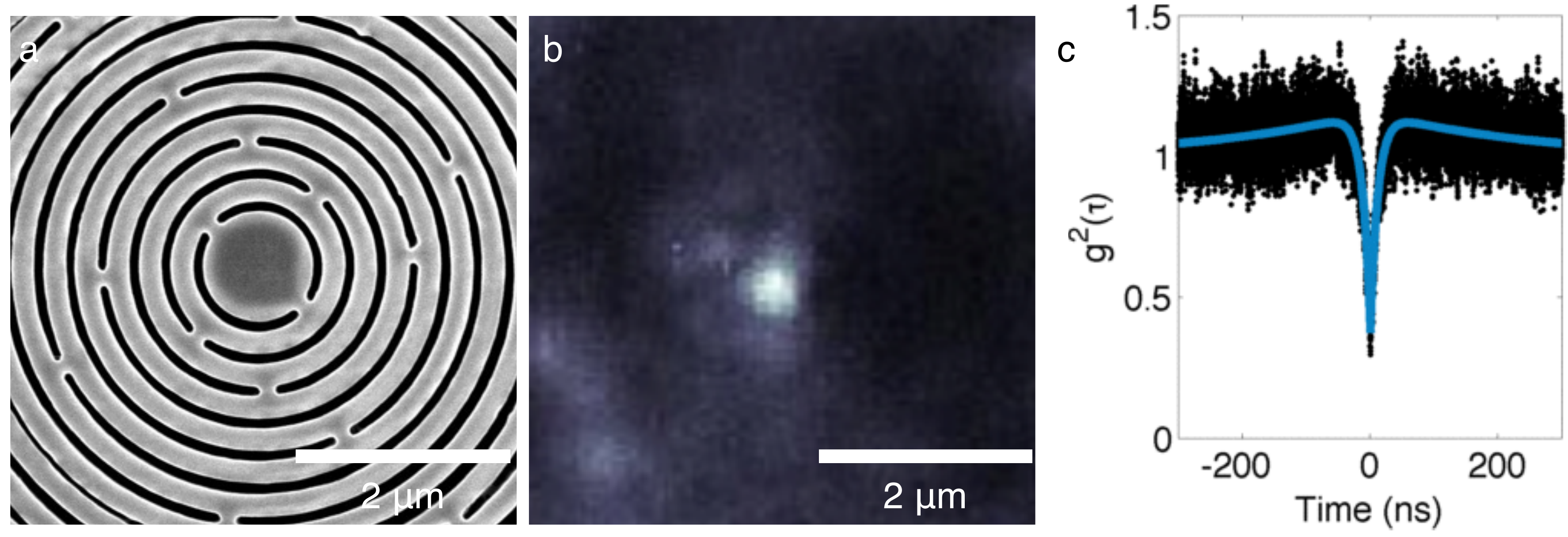}
	\caption{(a) Scanning electron micrograph and (b) PL scan of an NV within a diamond bullseye grating. (c) Normalized second-order auto-correlation measurement from the central bright spot in (b) with $g^{(2)}(0)=0.320\pm0.005$.}
\label{fig: 2}
\end{figure}

The bullseye gratings were investigated using a homebuilt confocal microscope with an oil objective (Nikon Plan Fluor NA$=1.3$). The PL scan in Fig.~2b shows a single NV inside a bullseye, as verified by a PL spectrum (see Supporting Information), and the anti-bunching in the auto-correlation histogram in Fig.~2c with $g^{(2)}(0) = 0.320\pm 0.005$.

We used a back-focal-plane (BFP) imaging technique to analyze the bullseye's far-field mode pattern. In a confocal imaging system, the Fourier transform of the far-field emission pattern is situated at the BFP of the objective lens. We imaged this onto a CCD camera (Princeton Instruments LN-1334) using a 400~mm lens (commonly called a `Bertrand lens'\cite{novotny2012principles}). The BFP image in Fig.~3d shows a strong intensity for modes of NA below 0.7, and a circular boundary for 1$<$NA$<$1.3. These results are consistent with the FDTD simulations (Fig.~3c) predicting that 13\% of the total emission occurs within an NA of 0.7 (See Supporting Information). In contrast, an NV in the unpatterned diamond membrane shows scattering primarily to high NA values, as seen in Fig.~3a (Simulation) and Fig.~3b (Experiment).

\begin{figure} 
	\centering
		\includegraphics[width=0.5\textwidth]{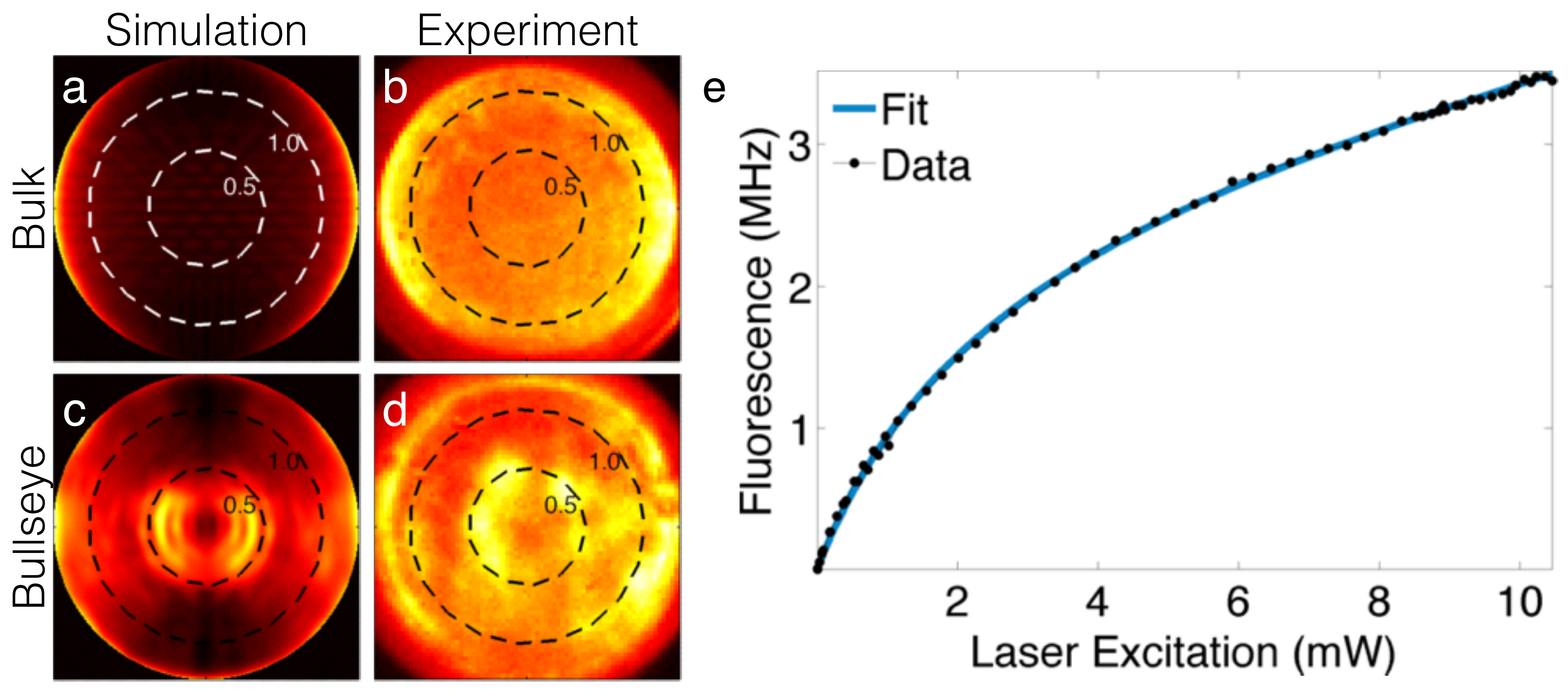}
	\caption{a-d: Simulated and experimental back focal plane images. The concentric circles are in units of numerical aperture, and the color intensities for each image is normalized to its respective maximum intensity value for wavelengths from 640-650~nm. Simulated far-field emission pattern of an NV oriented along a [111] diamond crystallographic direction inside a membrane \textit{with} (c) and \textit{without} (a) a grating structure. Measured far-field emission pattern of an NV in the $\sim$300~nm thick diamond membrane \textit{with} (d) and \textit{without} (b) a grating structure. (e) The saturation curve of the single NV indicates an asymptotic single photon count rate of $(3.0\pm0.1)\times10^6$~cts/s and a saturation excitation power of 2.5~mW.}
\label{fig: 3}
\end{figure}

We estimated the NV photon count rate from the PL saturation measurement shown in Fig.~3e. This curve is fit by a model of the PL rate of a single emitter with background fluorescence:

\begin{equation}
C(P) = \frac {C_{\infty}} {1+P_\text{sat}/P}+\alpha P,
\end{equation}

where $P$ is the excitation intensity, and the fit parameters are given by: $C_{\infty}$ the saturated single photon count rate, $P_\text{sat}$ the saturation excitation power, and $\alpha$ the linear background PL rate. We deduce a saturated count rate of $C_{\infty}=(3.0\pm0.1)\times10^6$~cts/s with a saturation power of 2.5~mW. This represents a 10-fold increase in count rate compared to what we observed for similarly deep NVs in bulk diamond. In a $\langle111\rangle$-oriented diamond substrate, one could expect a count rate increase of $\sim$30\%\cite{neu2014photonic} due to the improved alignment of the dipole with the plane of the bullseye.  

\begin{figure} 
	\centering
		\includegraphics[width=0.5\textwidth]{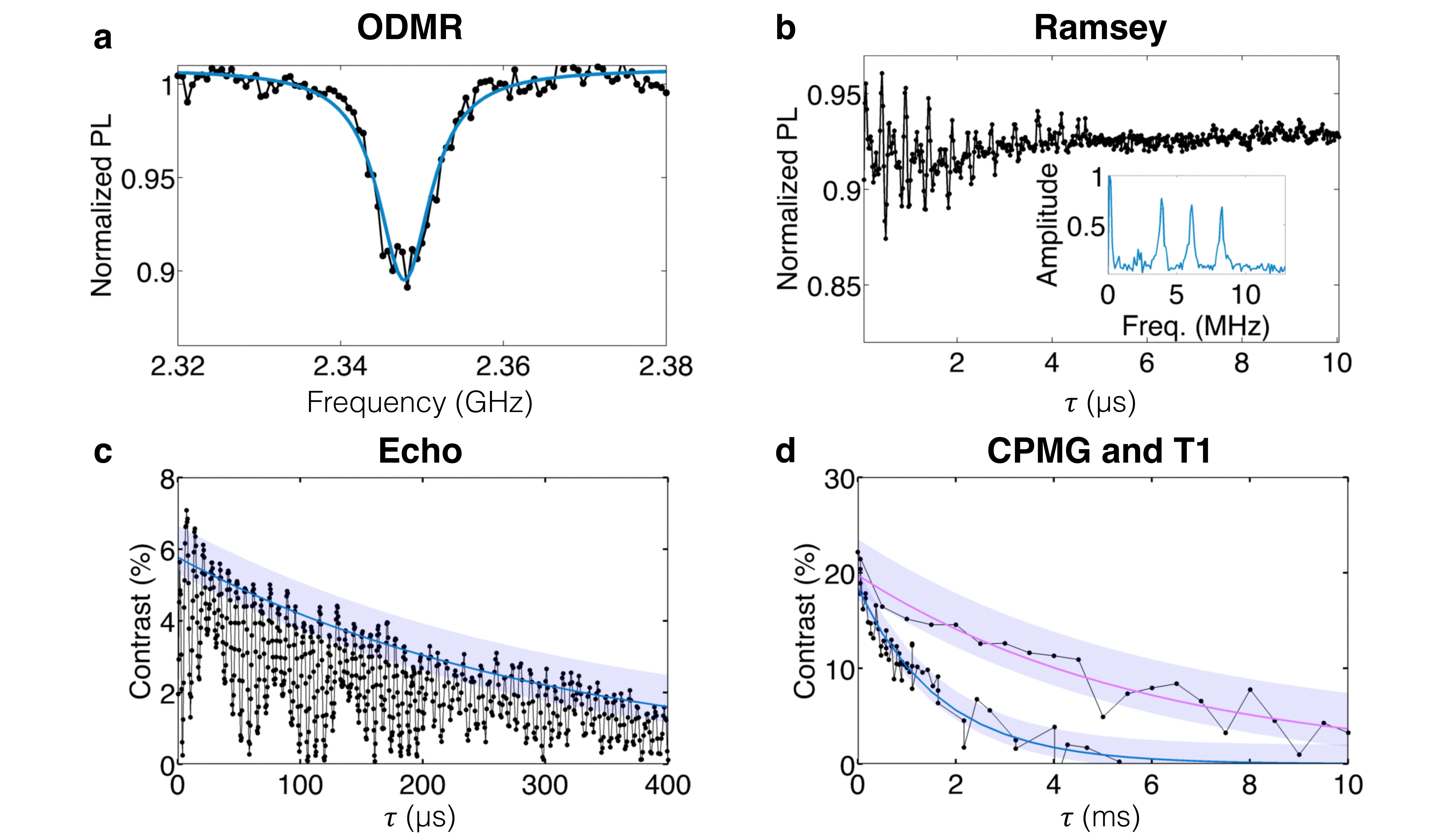}
	\caption{ Coherent spin control. (a) ODMR of the $m_s=0$ to $m_s=-1$ transition with an applied magnetic field of 18.6~mT along the NV-axis.  (b) A Ramsey sequence shows that the electronic transition is coupled to an $^{14}$N nuclear spin ($I=1$). The free-induction decay (FID) is limited by inhomogeneous broadening due to the surrounding spin bath of nuclear spins. \textit{Inset}: Fourier-transform of the FID signal showing three peaks corresponding to the hyperfine interaction with the nuclear spin. (c) The Hahn Echo sequence decouples the electronic spin with a quasi-static magnetic field, which extends the coherence time to $T_{2\text{,Hahn}}~=~311\pm23~\upmu$s. The shaded region around the exponential fit shows the 95\% confidence interval. (d) Further dynamical decoupling from the environment was achieved using a CPMG sequence yielding $T_{2\text{,CPMG}}=1.7\pm0.1$~ms (blue fit, black points). A single exponential fit indicates a spin-lattice relaxation time $T_1~=~5.9\pm0.5$~ms (purple fit, purple points).}
\label{fig: 4}
\end{figure}

NV centers inside the gratings exhibit spin coherence times similar to the parent CVD crystal. Fig.~4a shows optically-detected magnetic resonance (ODMR) under continuous optical and microwave excitation. Ramsey measurements in Fig.~4b indicate an $I=1$ spin of the host nitrogen, consistent with the expected $^{14}$N isotope for naturally occurring NVs. The phase coherence time ($T_{2\text{,Hahn}}$) was measured using a Hahn echo pulse sequence to cancel the dephasing by quasi-static magnetic fields\cite{2006.Science.Childress.coherent_nuclear_spin_dynamics}. From the exponential\cite{1367-2630-14-9-093004} decay envelope of the revivals in Fig.~4c, we determine $T_{2\text{,Hahn}}=311\pm23~\upmu$s. Carr-Purcell-Meiboom-Gill (CPMG) sequences further decoupled the NV spin and extended the coherence time through repeated spin-refocusing pulses. For a CPMG repetition order up to $n\sim150$ (See Supporting Information), we determine a $T_{2\text{,CPMG}}=1.7\pm0.1$~ms (Fig.~4d). Such $T_{2}$ values are typical for the parent diamond crystal, indicating that our nanofabrication process preserves the long electron spin coherence. 

Compared to other geometries with high collection efficiencies\cite{2009.NNano.Loncar.diamond_nanowire, pfaff2014unconditional}, the planar stucture of the bullseye grating allows for easy transfer onto different substrates for device integration with other optical components, such as on-chip photon detectors\cite{najafi2014chip,pernice_high-speed_2012,reithmaier_carrier_2014} and optical fiber facets\cite{schroder2010fiber}. As seen in the FDTD simulations in the Supporting Information, the bullseye structure shows a maximal collection efficiency of $\sim$30\% when the NV is located radially in the center of the bullseye. Simulations indicate that the collection efficiency remains within 50\% of the maximum even when the NV is within 10~nm of the diamond-air interface, which makes the bullseye structure attractive for sensing applications. For narrow-band applications ($\Delta\lambda/\lambda$<0.03) the collection efficiency can be optimized to as high as 90\% of the total dipole emission power within an NA=1.5 (See Supporting Information). This makes the bullseye geometry particularly useful for collection of the NV zero-phonon line, e.g. for spin-photon entanglement\cite{togan2010quantum, bernien2013heralded, pfaff2014unconditional}.

In summary, we demonstrate a nanophotonic device based on a circular `bullseye' grating to modify the broadband PL emission of a single NV center into smaller $k$-vectors in the far-field. The intrinsic coherence properties of the host materials were unaffected by the fabrication process, allowing for millisecond coherence times. The high collection efficiency provided by the bullseye structure promises improved proximal surface sensing \cite{grinolds2011quantum} and, combined with masked implantation\cite{toyli_chip-scale_2010}, allows for the scalable fabrication of high-performance quantum devices such as multi-qubit quantum network nodes \cite{dolde_room-temperature_2013,bernien2013heralded, pfaff2014unconditional,hausmann_coupling_2013}, room temperature single-photon sources for intensity standards\cite{aharonovich2011diamond}, and single-shot spin readout\cite{2010.Neumann-Jelezko-Wrachtrup.single-shot-readout,robledo2011high}.

\bibliography{references_bibdesk}

\section*{Acknowledgements}
Financial support was provided by the Air Force Office of Scientific Research PECASE (FA9550-13-1-0027) and the Quantum Memories MURI. E.H.C. was supported by the NASA Office of the Chief Technologist's Space Technology Research Fellowship. T.S. was supported in part by the Alexander von Humboldt Foundation. S.K. was supported in part by the MIT McGovern MINT program. Fabrication of devices was carried out in part at the Center for Functional Nanomaterials at Brookhaven National Laboratory, which is supported by the U.S. Department of Energy, Office of Basic Energy Sciences, under Contract No. DE-AC02-98CH10886. The authors thank Ming Lu and Matthew Trusheim for discussions, and Mircea Cotlet for help in optical measurements. 
\appendix
\onecolumngrid
\newpage
\section*{\selectfont\Large Supplementary Information}
\twocolumngrid
\begin{footnotesize}

\section{Spectrum of an NV inside the circular grating}

Fig.~1a shows an emission spectrum of an emitter at the center of the bullseye. The separation between peaks in the phonon-side band gives the free-spectral-range of a low-finesse ($F\sim1$) micro-cavity due to weak reflectance ($R\sim0.17$) at the gratings. This is in qualitative agreement with the expected spectrum (Fig.~1b), where we convolved a typical NV spectrum with the wavelength-dependent collection efficiency of an NV in a bullseye grating.

\begin{figure} 
	\centering
		\includegraphics[width=0.5\textwidth]{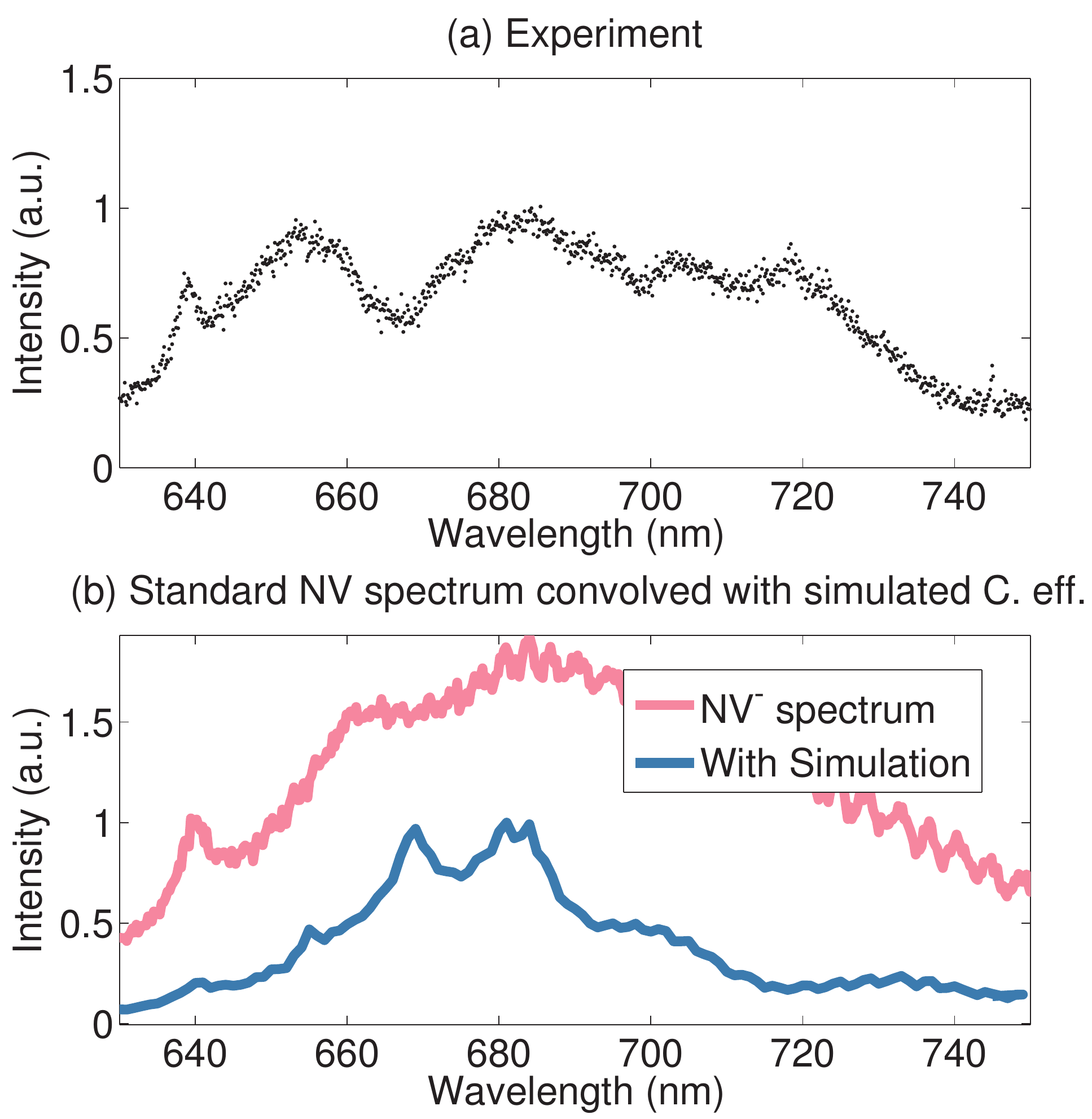}
	\caption{(a). Spectrum of an NV inside the bullseye grating. (b) Convolution of standard NV spectrum (pink) with a simulated, wavelength-dependent collection efficiency (blue).}
\label{fig: 1}
\end{figure}

\section{Collection Efficiency of NV in bullseye}

To investigate the relationship between NV depth and the collection efficiency of the bullseye, the dipole emitter in the simulation is scanned along the $z$-axis of the bullseye structure. The collection efficiency calculated from 3D-FDTD simulations by projecting the electric-field intensity to the far-field is shown in Fig.~2.

In the context of sensing applications, Fig.~3 shows a simulated wavelength-dependent collection efficiency within a collection aperture of 1.5 for the grating design used in the main text. The simulation shows that it is possible to achieve above 90\% of the collection efficiency from some wavelength range by using a bullseye grating. 

For spin-photon entanglement applications, a high collection efficiency with low-NA emission can be achieved for a spectrally-resolved region around the NV's zero-phonon line (637~nm). Simulated images in Fig.~4 show that a low-NA emission with high collection efficiency (40\%) is possible with bullseye gratings, and would be useful for cryogenic optical experiments which typically require long working-distance, low-NA objectives.
\begin{figure} 
	\centering
		\includegraphics[width=0.5\textwidth]{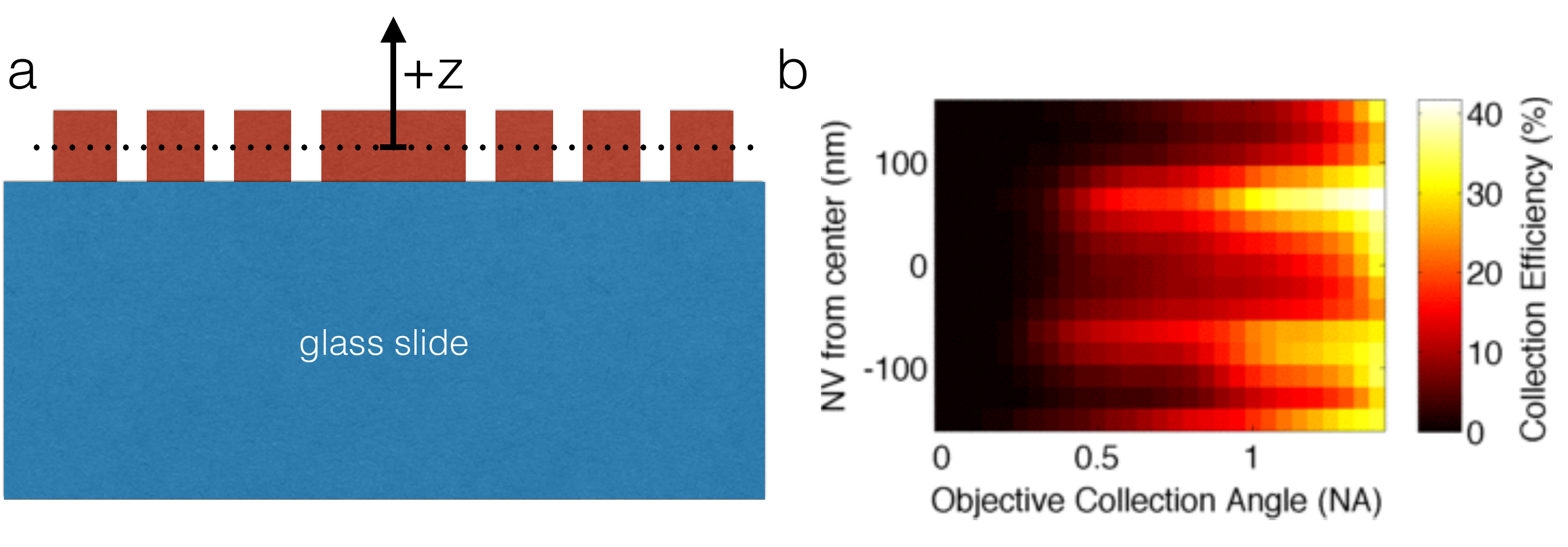}
	\caption{(a) A sketch of the geometry denoting $z$, the axis along which the dipole was placed for the simulation found in (b) -- in which we show the collection efficiency of NV in bullseye as a function of objective NA and distance from the center of a $\sim$300~nm diamond membrane.}
\label{fig: 2}
\end{figure}

\begin{figure} 
	\centering
		\includegraphics[width=0.5\textwidth]{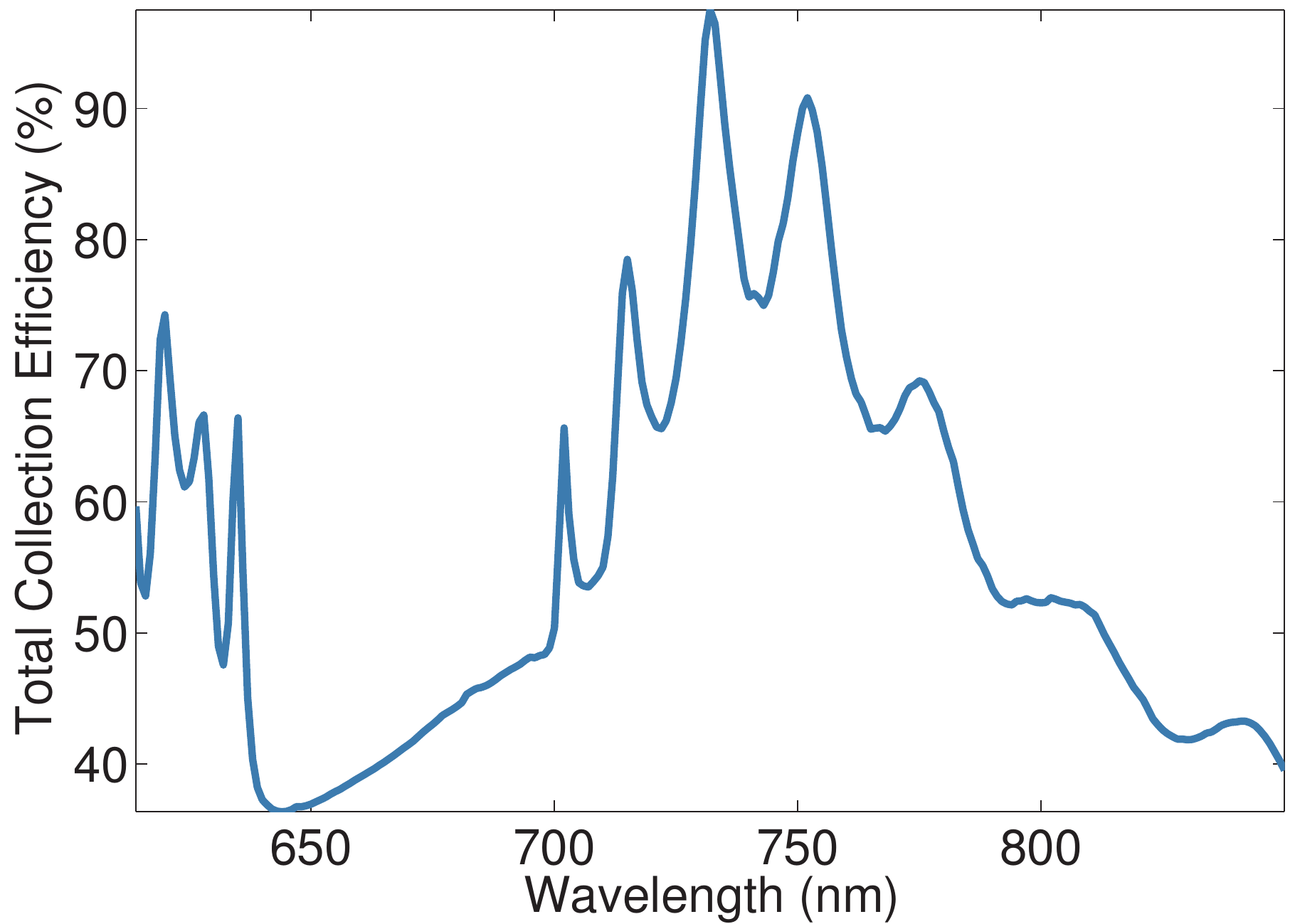}
	\caption{A simulated plot of the total collection efficiency in the downwards direction within a collection aperture of 1.5.}
\label{fig: 2}
\end{figure}

\begin{figure} 
	\centering
		\includegraphics[width=0.5\textwidth]{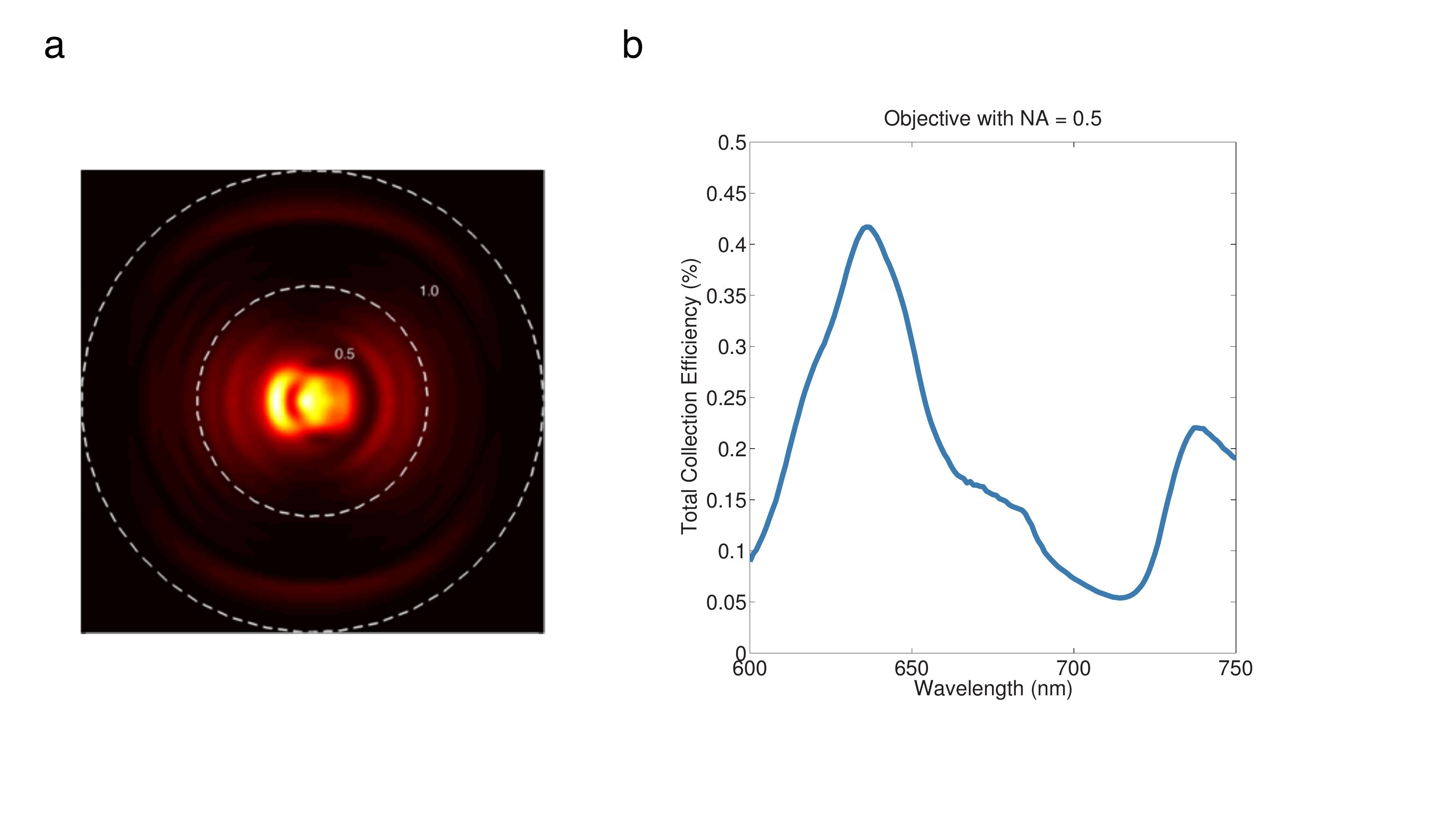}
	\caption{(a) A plot of the simulated emission of $637$~nm into low NA-modes from an optimized bullseye structure. The concentric dotted circles have units of NA. (b) Total collection efficiency with an NA = 0.5 objective from an bullseye grating optimized for the NV zero-phonon line at 637~nm.}
\label{fig: 2}
\end{figure}

\section{Spin Measurements}

\begin{figure} 
	\centering
		\includegraphics[width=0.5\textwidth]{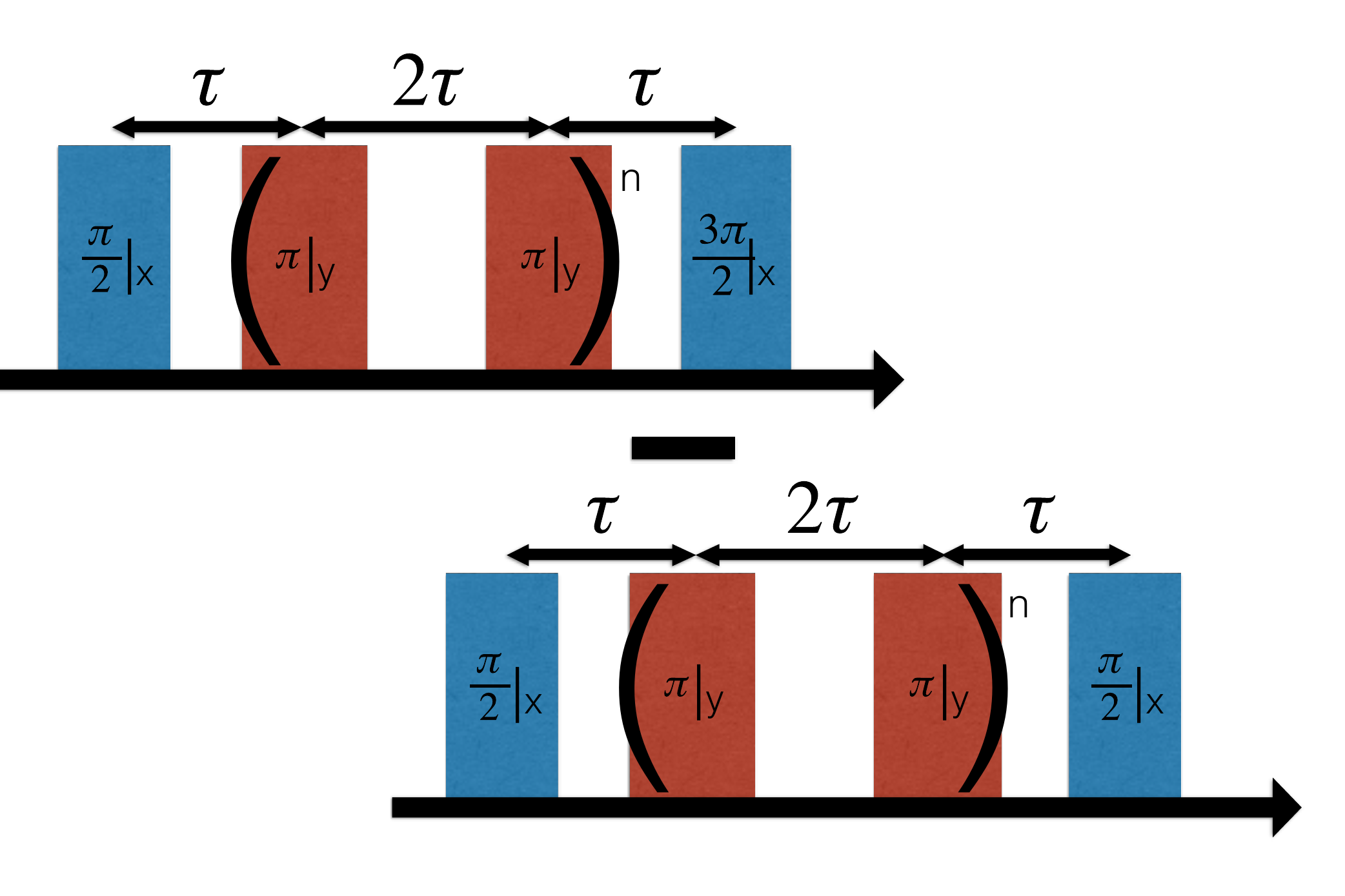}
	\caption{Pulse sequence for CPMG measurement and normalization.}
\label{fig: 2}
\end{figure}

For determining the T$_{2,CPMG}$ of an NV using CPMG, we sampled values of $\tau$ only at the revivals measured from a standard Hahn echo measurement, $\left(\frac{\pi}{2}\right)_x-\left(\pi\right)_x-\left(\frac{\pi}{2}\right)_x$. The CPMG decay was sampled by keeping a constant delay between the $\pi$-pulses while increasing the number ($n$) of $\pi$-pulses (Fig.~5). The signal was normalized to a $\left(\frac{\pi}{2}\right)_{-x}$ rotation, or equivalently a $\left(\frac{3\pi}{2}\right)_x$ rotation. Such a normalization method is equivalent to projecting the measurement basis either onto the m$_s$=0 basis (bright, $\left(\frac{3\pi}{2}\right)_x$), or the m$_s$=1 basis (dark, $\left(\frac{\pi}{2}\right)_x$). This method of sampling allows for a sampling of the decoherence\cite{naydenov_dynamical_2011}.

\end{footnotesize}

\end{document}